\documentclass[aps,prl,twocolumn,showpacs,superscriptaddress]{revtex4}
\usepackage{graphicx}

\begin{document}

\title{Antiferromagnetic order and superlattice structure in nonsuperconducting and superconducting Rb$_y$Fe$_{1.6+x}$Se$_2$}
\author{Meng Wang}
\affiliation{Beijing National Laboratory for Condensed Matter Physics, Institute of Physics, Chinese Academy of Sciences, Beijing 100190, China}
\affiliation{Department of Physics and Astronomy, The University of Tennessee, Knoxville, Tennessee 37996-1200, USA }
\author{Miaoyin Wang}
\affiliation{Department of Physics and Astronomy, The University of Tennessee, Knoxville, Tennessee 37996-1200, USA }
\author{G. N. Li}
\affiliation{Beijing National Laboratory for Condensed Matter Physics, Institute of Physics, Chinese Academy of Sciences, Beijing 100190, China}
\affiliation{NIST Center for Neutron Research, National Institute of Standards and Technology, Gaithersburg, MD 20899, USA}
\author{Q. Huang}
\affiliation{NIST Center for Neutron Research, National Institute of Standards and Technology, Gaithersburg, MD 20899, USA}
\author{C. H. Li}
\affiliation{Beijing National Laboratory for Condensed Matter Physics, Institute of Physics, Chinese Academy of Sciences, Beijing 100190, China}
\author{G. T. Tan}
\affiliation{Department of Physics and Astronomy, The University of Tennessee, Knoxville, Tennessee 37996-1200, USA }
\affiliation{College of Nuclear Science and Technology, Beijing Normal University, Beijing 100875, China}
\author{C. L. Zhang}
\affiliation{Department of Physics and Astronomy, The University of Tennessee, Knoxville, Tennessee 37996-1200, USA }
\author{Huibo Cao}
\affiliation{Neutron Scattering Science Division, Oak Ridge National Laboratory, Oak Ridge, Tennessee 37831-6393, USA}
\author{Wei Tian}
\affiliation{Ames Laboratory and Department of Physics and Astronomy, Iowa State University, Ames, Iowa 50011, USA}
\author{Yang Zhao}
\affiliation{NIST Center for Neutron Research, National Institute of Standards and Technology, Gaithersburg, MD 20899, USA}
\affiliation{Department of Materials Science and Engineering, University of Maryland, College Park, MD 20742, USA}
\author{Y. C. Chen}
\affiliation{Beijing National Laboratory for Condensed Matter Physics, Institute of Physics, Chinese Academy of Sciences, Beijing 100190, China}
\author{X. Y. Lu}
\affiliation{Beijing National Laboratory for Condensed Matter Physics, Institute of Physics, Chinese Academy of Sciences, Beijing 100190, China}
\author{Bin Sheng}
\affiliation{Beijing National Laboratory for Condensed Matter Physics, Institute of Physics, Chinese Academy of Sciences, Beijing 100190, China}
\author{H. Q. Luo}
\affiliation{Beijing National Laboratory for Condensed Matter Physics, Institute of Physics, Chinese Academy of Sciences, Beijing 100190, China}
\author{S. L. Li}
\affiliation{Beijing National Laboratory for Condensed Matter Physics, Institute of Physics, Chinese Academy of Sciences, Beijing 100190, China}
\author{M. H. Fang}
\affiliation{Department of Physics, Zhejiang University, Hangzhou 310027, China}
\author{J. L. Zarestky }
\affiliation{Ames Laboratory and Department of Physics and Astronomy, Iowa State University, Ames, Iowa 50011, USA}
\author{W. Ratcliff}
\affiliation{NIST Center for Neutron Research, National Institute of Standards and Technology, Gaithersburg, MD 20899, USA}
\author{M. D. Lumsden}
\affiliation{Neutron Scattering Science Division, Oak Ridge National Laboratory, Oak Ridge, Tennessee 37831-6393, USA}
\author{J. W. Lynn}
\affiliation{NIST Center for Neutron Research, National Institute of Standards and Technology, Gaithersburg, MD 20899, USA}
\author{Pengcheng Dai}
\email{pdai@utk.edu}
\affiliation{Department of Physics and Astronomy, The University of Tennessee, Knoxville,
Tennessee 37996-1200, USA }
\affiliation{Beijing National Laboratory for Condensed Matter Physics, Institute of Physics, Chinese Academy of Sciences, Beijing 100190, China}

\begin{abstract}
Neutron diffraction has been used to study the lattice and magnetic structures of the insulating and superconducting Rb$_y$Fe$_{1.6+x}$Se$_2$.  For the insulating Rb$_{y}$Fe$_{1.6+x}$Se$_2$, neutron polarization analysis and single crystal neutron diffraction unambiguously confirm
the earlier proposed $\sqrt{5}\times\sqrt{5}$ block antiferromagnetic structure.  For superconducting 
samples ($T_c=30$ K), we find that in addition to the tetragonal $\sqrt{5}\times\sqrt{5}$ superlattice structure transition at 513 K, the material develops a separate $\sqrt{2}\times \sqrt{2}$ superlattice structure at a lower temperature of 480 K.
These results suggest that superconducting Rb$_{y}$Fe$_{1.6+x}$Se$_2$ is phase separated with coexisting 
$\sqrt{2}\times \sqrt{2}$ and $\sqrt{5}\times\sqrt{5}$ superlattice structures.
\end{abstract}


\maketitle

\section{I. Introduction}
The recent discovery of superconductivity around 30 K in alkaline iron selenides $A_y$Fe$_{1.6+x}$Se$_2$  ($A = $ K, Rb, Cs, Tl) \cite{jgguo,krzton,mhfang,afwang} has generated considerable excitement in the condensed matter physics community because the parent compounds of these materials are antiferromagnetic (AFM) insulators \cite{mhfang,haggstrom} instead of being AFM metals as the iron arsenide superconductors \cite{kamihara,cruz}.  Because of their metallic nature, band structure calculations for iron arsenides have predicted the presence of the hole-like Fermi surfaces at the  $(0,0)$ point and electron-like Fermi surfaces at the $M(\pi,0)/(0,\pi)$ points in the 
Brioullion zone using an orthorhombic full lattice unit cell \cite{johnston,mazin}.  
As a consequence, Fermi surface nesting and quasiparticle excitations between the hole and electron pockets can give rise to static AFM spin-density-wave order at the in-plane wave vector $Q = (\pi,0)$ \cite{jdong}.  Indeed, neutron diffraction experiments have confirmed the 
$Q =( \pi,0)$ AFM order in the parent compounds of iron arsenide superconductors, and doping to induce superconductivity suppresses the
static AFM order \cite{cruz}.   In addition, angle resolved photoemission (ARPES) measurements \cite{hding} have identified the expected hole and electron pockets in superconducting iron arsenides, thus providing evidence for superconductivity arising from the sign reversed electron-hole inter-pocket excitations \cite{mazin,kuroki,chubkov,fwang}.  

If Fermi surface nesting and electron-hole pocket excitations are essential ingredients for magnetism and superconductivity in Fe-based superconductors \cite{cruz,mazin,jdong,kuroki,chubkov,fwang}, alkaline iron selenide superconductors should behave differently from iron arsenides since ARPES measurements on these materials reveal only electron Fermi surfaces at $M(\pi,0)/(0,\pi)$ points 
and no hole Fermi surface at  $(0,0)$ point \cite{yzhang,xpwang,xjzhou}.  Indeed, recent transmission electron microscopy \cite{jqli}, X-ray and neutron diffraction experiments \cite{zavalij,bacsa,wbao1,pomjakushin1,fye,wbao2,pomjakushin} have confirmed that the Fe vacancies in 
$A_y$Fe$_{1.6+x}$Se$_2$ form a $\sqrt{5}\times\sqrt{5}$ superlattice order as shown in Fig. 1(a) \cite{haggstrom}.   Furthermore, a block-type AFM structure with a large moment aligned along the $c$-axis [Figs. (1a) and (1b)] has been proposed for both superconducting and insulating  
$A_y$Fe$_{1.6+x}$Se$_2$ based on Rietveld analysis of neutron powder diffraction data \cite{wbao1,wbao2}.  In stark contrast to other Fe-based superconductors, where optimal superconductivity generally 
occurs in the absence of a static AFM order \cite{jzhao1}, the large moment AFM order is believed to co-exist with superconductivity microscopically \cite{shermadini} and the superconducting phase develops without much affecting the AFM order \cite{wbao2}.
If magnetic moments up to 3.3 $\mu_B$ per Fe  indeed coexist with optimal superconductivity microscopically 
in $A_y$Fe$_{1.6+x}$Se$_2$ as suggested in powder neutron diffraction \cite{wbao1,wbao2} and muon rotation experiments \cite{shermadini}, the electronic phase diagram in this class of materials will be much different than the other Fe-based superconductors \cite{johnston}.  Since these new materials pose a major challenge to the current theories of superconductivity \cite{mazin2011}, it is important to confirm the proposed magnetic structure in single crystals and determine its relationship with superconductivity.

In this article, we present comprehensive neutron diffraction measurements on 
powder and single crystals of nonsuperconducting and superconducting Rb$_{y}$Fe$_{1.6+x}$Se$_{2}$.
We used neutron polarization analysis to separate the magnetic from nuclear scattering.  From the Rietveld analysis of the neutron powder diffraction data on nonsuperconducting
 Rb$_{0.89}$Fe$_{1.58}$Se$_2$ \cite{mywang2011}, we confirm the previously reported $\sqrt{5}\times\sqrt{5}$
Fe vacancy order with $I4/m$ space group \cite{haggstrom}.  
Since Rietveld analysis of the powder diffraction pattern cannot conclusively separate the proposed block  
AFM structure from the quaternary collinear AFM structure with the $I112^\prime/m^\prime$ space group \cite{wbao1,pomjakushin1,fye,wbao2,pomjakushin}, we used four circle 
single crystal diffractometer to measure Bragg peaks associated with each AFM structure, and confirmed 
the proposed block AFM structure \cite{wbao1}.  
For superconducting Rb$_{0.75}$Fe$_{1.63}$Se$_{2}$ ($Tc=30$ K), we find that in addition to the
$\sqrt{5}\times\sqrt{5}$ block AFM structure, the sample exhibits 
 a quasi-two-dimensional $\sqrt{2}\times \sqrt{2}$  
superlattice distortion associated with wave vectors $Q=(0.5,0.5,L)$, where $L=$ integers.  
These results suggest that lattice structures in 
superconducting Rb$_{y}$Fe$_{1.6+x}$Se$_{2}$ are more complicated than the pure 
$\sqrt{5}\times\sqrt{5}$ superlattice unit cell, consistent with nanoscale phase separation seen by
transmission electron microscopy \cite{jqli1,jqli2} and 
X-ray diffraction experiments \cite{ricc1,ricc2}.

\begin{figure}[h]
\includegraphics[scale=.38]{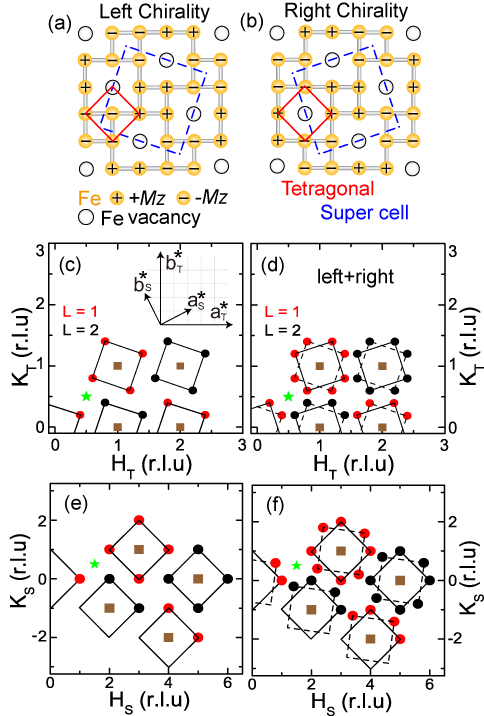}
\caption{(color online) (a) Schematic of Fe nuclear and spin structures in left chirality and (b) in right chirality. The plus and minus signs represent spin directions up and down along the $c$-axis, respectively. The solid and dashed squares are the top views of 
the tetragonal cell (red) and the $\sqrt{5}\times\sqrt{5}$  superlattice cell (blue), respectively. 
(c) The expected Bragg peak positions in the $[H_T, K_T]$ tetragonal unit cell notation for the left chirality. The red circles with $L=1$ are magnetic peaks and the black circles are nuclear peaks with $L=2$.  The brown squares are nuclear peaks for both even and odd $L$. The green star at (0.5,0.5) rlu 
is not an expected Bragg peak position for the present structure.  
The inset is a schematic of the tetragonal and superlattice unit cell in reciprocal space.
(d) The expected Bragg peaks from both left and right chiralities.
(e,f) The same expected Bragg peaks in $[H_S, K_S]$ superlattice unit cell notation.
}
\end{figure}

\section{II. Experimental Details}

We have carried out neutron diffraction experiments at the BT-1 powder diffractometer 
and BT-7 thermal triple-axis spectrometer at the National Institute for Standard and Technology Center for 
Neutron Research.  We have also performed additional measurements at HB-1A triple-axis spectrometer and HB-3A four circle single crystal diffractometer at the High-Flux Isotope Reactor, Oak Ridge National Laboratory.  Our experimental setup for the BT-1 powder diffraction measurements was described previously \cite{cruz}.  For BT-7 measurements, we used polarized neutron scattering to separate the magnetic from nonmagnetic scattering processes \cite{jwlynn,lipscombe}. 
In previous powder diffraction measurements on $A_y$Fe$_{1.6+x}$Se$_2$ near $x=0$ \cite{wbao1,wbao2,pomjakushin1,pomjakushin}, the iron atoms were found to form an ordered vacancy structure with a $\sqrt{5}\times\sqrt{5}\times1$ superlattice unit cell.  Although a block AFM spin structure with space group $I4/m^\prime$ [Fig. 1(a) and Fig. 1(b)] 
 was identified \cite{wbao1,wbao2}, powder Rietveld analysis cannot conclusively distinguish the block AFM structure from a stripe-like AFM structure with $I112^\prime/m^\prime$ space group [Fig. 4(b)] \cite{wbao2,pomjakushin}.  We have therefore used the HB-3A single crystal diffractometer to 
 measure all the accessible Bragg peaks, and including the nonequivalent magnetic reflections with the same 
 momentum transfer that are fully overlapped in the powder diffraction experiments, thus
providing more information to separate these two magnetic structures.
HB-3A uses a vertically focusing Si(2,2,0) monochromator with fixed wavelength of 1.536 \AA\ \cite{chako}.  The HB-1A 
 triple-axis spectrometer has horizontal collimation $48^\prime-48^\prime-40^\prime-68^\prime$ with fixed incident beam energy of $E_i=14.7$ meV.

\begin{figure}[h]
\includegraphics[scale=.4]{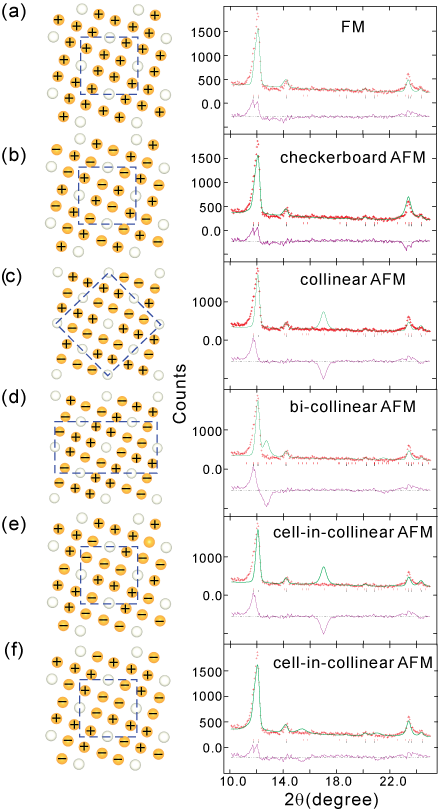}
\caption{(color online) Portion of the neutron powder diffraction pattern for Rb$_{0.89}$Fe$_{1.58}$Se$_2$ at 480 K and
its comparison with the expected neutron powder patterns for various proposed magnetic structures 
of $A_y$Fe$_{1.6+x}$Se$_2$ \cite{xwyan}.  The plus and minus signs denote spin directions parallel and antiparallel to the $c$-axis, respectively.  The data and model fitted difference plots are shown below.  The
weighted $R$ factors for each model are listed as follows: 
(a) Ferromagnetic or FM. $R_{wp}=7.81$\%. (b) checkerboard AFM. $R_{wp}=7.56$\%. 
(c) collinear AFM. $R_{wp}=9.93$\%. (d) bi-collinear AFM. $R_{wp}=9.7$\%. 
(e) Cell-in-collinear AFM. $R_{wp}=7.45$\%.  (f) Cell-in-collinear AFM. $R_{wp}=7.44$\%.
None of the models can fit the observed neutron powder diffraction pattern.
}
\end{figure}

We begin our discussion by specifying the real and reciprocal space notations used in this article. 
Figures 1(a) and 1(b) show the left and right chiralities of the proposed block AFM structures, respectively. 
The blue dashed lines show the structural and 
magnetic unit cells for the $\sqrt{5}\times\sqrt{5}$ Fe vacancy structure, while the red
solid lines are the $I4/mmm$ symmetry tetragonal unit cell suitable 
for doped BaFe$_2$As$_2$ \cite{johnston}. The $+,-$ signs indicate the Fe moment directions parallel and anti-parallel to the $c$-axis, respectively.
For easy comparison with previous work in iron pnictides, we define  
wave vector $Q=(q_x,q_y,q_z)$ in \AA$^{-1}$ as $(H_T;K_T;L_T)=(q_xa_T/2\pi;q_yb_T/2\pi;q_zc_T/2\pi)$ reciprocal lattice units (rlu),
where $a_T=b_T\approx 3.9$ \AA\ are lattice parameters for tetragonal unit cell of iron pnictides \cite{lynn}. The Bragg peaks in the $\sqrt{5}\times\sqrt{5}$ superlattice unit cell can be indexed as $(H_S;K_S;L_S)=(q_xa_S/2\pi;q_yb_S/2\pi;q_zc_S/2\pi)$ rlu, where $a_S=b_S=\sqrt{5}\times a_T=\sqrt{5}\times b_T=8.73$ \AA\ and $c_S=c_T=14.11$ \AA\ for the nonsuperconducting
Rb$_{0.89}$Fe$_{1.58}$Se$_2$, $a_S=b_S=\sqrt{5}\times a_T=\sqrt{5}\times b_T=8.74$ \AA\ and $c_S=c_T=14.47$ \AA\ for the 
superconducting Rb$_{0.75}$Fe$_{1.63}$Se$_{2}$.
The Rb, Fe, and Se compositions are determined from inductively coupled plasma atomic emission spectroscopy analysis.

\begin{figure}[h]
\includegraphics[scale=.4]{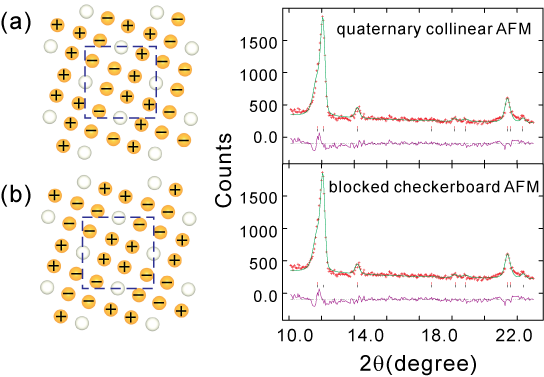}
\caption{(color online) Two possible AFM structures that can equally well fit the neutron powder diffraction pattern of Rb$_{0.89}$Fe$_{1.58}$Se$_2$ at 480 K.  (a) Quaternary collinear AFM structure and (b) block AFM structure.  The $R_{wp}$ factors
for these two magnetic structures are 7.0\% and 6.95\%, respectively.  
Both of these models were proposed earlier \cite{wbao2,pomjakushin}.
}
\end{figure}

In iron pnictides, the Fe moments are in the Fe plane along the $a$-axis direction \cite{lynn}. The magnetic 
Bragg peaks occur at $[m\pm 0.5,n\pm 0.5,L]_T$ ($m,n=0,\pm1, \pm2, \pm 3...,L=odd$) positions in tetragonal unit cell notation. 
For $A_y$Fe$_{1.6+x}$Se$_2$ with the block AFM structure in Figure 1, the magnetic peaks from left chirality are expected at $(H,K,L)_T=(0.2+2m+\delta,-0.4+n,L_T);(0.4+2m+\delta,0.2+n,L_T);(-0.2+2m+\delta,0.4+n,L_T);(-0.4+2m+\delta,-0.2+n,L_T), (m,n=0,\pm1,\pm2,...,L_T=\pm1,\pm3,\pm5...,$ when $n$ is even, $\delta=0;$ $n$ is odd, $\delta=1)$ as shown in Figs. 1(c) and 1(e). 
If one considers both left and right chiralities, 
the magnetic peaks will double and occur at $(H,K,L)_T=(\pm0.2+2m+\delta,\pm0.4+n,L_T);(\pm0.4+2m+\delta,\pm0.2+n,L_T), (m,n=0,\pm1,\pm2,...,L_T=\pm1,\pm3,\pm5...,$ when $n$ is even, $\delta=0;$ $n$ is odd, $\delta=1)$. 
The nuclear Bragg reflections can be indexed the same way, but with 
 $L=even$ and when $n$ is even, $\delta=1$; $n$ is odd, $\delta=0$. 
The squares in Figs. 1(c)-1(f) indicate nuclear Bragg peak positions in tetragonal and superlattice unit cell notation.
The conversion of Miller indices between tetragonal and superlattice unit cell for left chirality 
is as follows:

\[\left(\begin{array}{clcr}H_S\\K_S\end{array}\right)=
\left(\begin{array}{clcr}
2&1  \\
-1&2
\end{array}\right)
\left(\begin{array}{clcr}H_T\\K_T\end{array}\right)
\] 

For right chirality, the conversion is:

\[\left(\begin{array}{clcr}H_S\\K_S\end{array}\right)=
\left(\begin{array}{clcr}
2&-1  \\
1&2
\end{array}\right)
\left(\begin{array}{clcr}H_T\\K_T\end{array}\right).
\]

Our single crystals of Rb$_y$Fe$_{1.6+x}$Se$_2$ were grown using the Bridgeman method. 
First, Fe$_{2+\delta}$Se$_2$ 
powders were prepared with a high-purity powder of selenium
(Alfa Aesar, 99.99\%) and iron (Alfa Aesar, 99.9\%) as described in Ref. \cite{bacsa}. The Fe$_{2+\delta}$Se$_2$ and Rb (Alfa
Aesar, 99.75\%) were then mixed in appropriate stoichiometry and 
were put into an alumina crucible.  The crucible was sealed in an evacuated
silica ampoule. The mixture was heated up to 1030$^\circ$C and kept
over 3 h.  Afterward the melt was cooled down to 730$^\circ$C with
a cooling rate of 6$^\circ$C/h, and finally the furnace was cooled
to room temperature with the power shut off. Well-formed
black crystals were obtained which could be easily cleaved
into plates with flat shiny surfaces.
We have also grown Rb$_y$Fe$_{1.6+x}$Se$_2$ single crystals using flux method as described in
Ref. \cite{afwang}.

For BT-1 powder diffraction measurements, 
we ground $\sim 2$ grams of single crystals of Rb$_{0.89}$Fe$_{1.58}$Se$_2$ and Rb$_{0.75}$Fe$_{1.63}$Se$_{2}$.
For the experiment on HB-3A, we used a $\sim$0.5 gram single crystal from the same batch of Rb$_{0.89}$Fe$_{1.58}$Se$_2$. 
For HB-1A unpolarized and BT-7 polarized neutron scattering measurements, we used 
$~0.8$ gram single crystals 
of  Rb$_{0.89}$Fe$_{1.58}$Se$_2$ and Rb$_{0.75}$Fe$_{1.63}$Se$_{2}$
with less than 1$^\circ$ mosaic aligned in the $[H,H,L]_T$ zone in tetragonal notation. 
To separate the magnetic order from nonmagnetic scattering processes, we performed neutron polarization analysis, where 
the neutron spin flip (SF) scattering for  
polarization direction parallel to the scattering plane (HF) gives pure magnetic scattering \cite{lipscombe}. In the BT-7 setup, the spin polarization direction in the incident beam could be changed via a flipper and the 
spin polarization direction for the scattered beam was fixed.  The neutron SF magnetic scattering 
corresponds to flipper on, while the nuclear coherent 
scattering is with flipper off, which corresponds to nonspin flip (NSF) scattering.
A horizontal guide field was directed along the in-plane momentum transfer (HF configuration), and the
flipping ratio of $\sim$22 was obtained in this HF field configuration using an incident energy of 14.7 meV. 
A pyrolytic graphite (PG) filter was placed in the incident beam direction to suppress $\lambda/2$ scattering.
A position-sensitive detector (PSD) was used with $open-80^\prime-80^\prime-radial$ collimations. 
For all the polarized measurements,
the sample was at room temperature 
and aligned in both the $[H,H,L]_T$ and $[H,0,L]_T$ zones to reach the desired reciprocal space by tilting 
the sample goniometer.

\section{III. Results}

We first discuss our neutron powder refinement results on Rb$_{0.89}$Fe$_{1.58}$Se$_2$ with the goal of determining the 
magnetic structure of the system.  
In previous theoretical work \cite{xwyan}, eight possible magnetic structures have been proposed for the $\sqrt{5}\times\sqrt{5}$ 
iron vacancy superlattice unit cell.  Figure 2 summarizes the comparison between the observed neutron diffraction intensity and calculated intensity for six suggested magnetic structures.  As we can see from the figure, all six magnetic models fail to describe the observed spectrum.  

\begin{figure}[t]
\includegraphics[scale=.45]{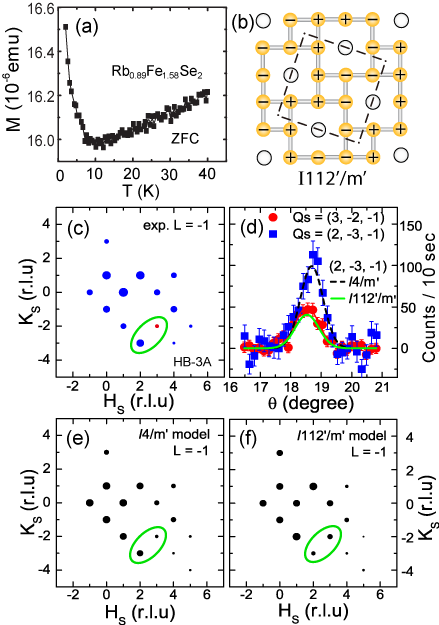}
\caption{(color online) (a) Zero field cooled (ZFC) magnetic susceptibility of 17.12 mg Rb$_{0.89}$Fe$_{1.58}$Se$_2$ at 30 Oe with $H//ab$ plane measured using a commercial SQUID. No superconductivity was observed.
(b) Magnetic structure in the $I112^\prime/m^\prime$ space group representation. 
(c) The experimental data collected from HB-3A.  
The data plotted are from one chiral domain. 
In single crystal diffraction experiments, we compared several peaks from lef and right chiral domains
and found them to be the same. We used one domain for the data collection, 
since this can save half of the beam time for a point detector diffractometer.
The radius of the circles are proportional to the intensity of the Bragg peaks. 
The two peaks enclosed 
in the green ellipse have quite different intensities for the $I4/m^\prime$ and $I112^\prime/m^\prime$ AFM models.
(d) The rocking curves for Bragg peaks $Q_S=(3,-2,-1)$ (red circle) and $Q_S=(2,-3,-1)$ (blue square). The black and green solid lines are the simulation of the expected magnetic intensities for $(2,-3,-1)$ in $I4/m^\prime$ and $I112^\prime/m^\prime$ models, respectively, where the intensity of the $(3,-2,-1)$ peak 
are normalized to the value of the experiment. Error bars represent one standard deviation. 
(e) The simulations of the expected Bragg peak intensity with the $I4/m^\prime$ model.
(f) Identical simulation with the $I112^\prime/m^\prime$ model.      
}
\end{figure}

In previous neutron powder diffraction work \cite{wbao1,pomjakushin1,fye,wbao2,pomjakushin}, it has been suggested that the block AFM structure Fig. 3(b) and 
the quaternary collinear
AFM structure in Fig. 3(a) can both fit the observed neutron diffraction spectra \cite{wbao2,pomjakushin}.  
Our Rietveld analysis on  Rb$_{0.89}$Fe$_{1.58}$Se$_2$ for both AFM structures shown in Fig. 3 confirms this result.
Although the block AFM structure is thought to be more energetically favorable \cite{wbao2}, 
the AFM structure shown in Fig. 3(a) is not conclusively ruled out \cite{pomjakushin}.

\begin{figure}[t]
\includegraphics[scale=.5]{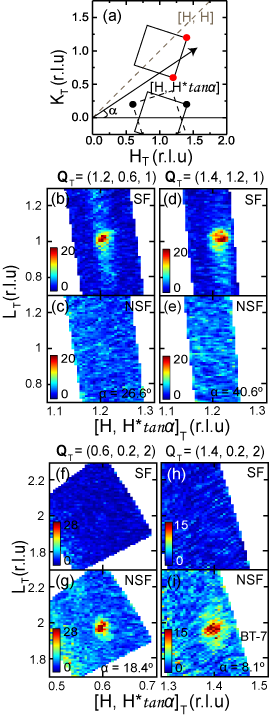}
\caption{(color online) Neutron polarization analysis with neutron guide field in the horizontal scattering plane along the wave vector direction for Rb$_{0.89}$Fe$_{1.58}$Se$_2$. 
(a) Schematics of the reciprocal space probed in the $[H_T,K_T]$ plane. 
The scans are along the arrow direction which depends on the lower goniometer arc tilting angles $\alpha$.
The observed scattering intensity in SF and NSF channels are plotted in 
 the $[H,Htan\alpha,L]_T$ plane of reciprocal space. 
(b) Magnetic peak $Q_T=(1.2,0.6,1)$ appears in the SF (flipper on) channel and disappears in the 
(c) NSF channel. (d),(e) Magnetic peak $Q_T=(1.4,1.2,1)$ in the SF and NSF channels, respectively.
(f-i) Nuclear peaks $Q_T=(0.6,0.2,2)$ and $Q_T=(1.4,0.2,2)$ show no observable scattering in the SF channel and
the entire peak appears in the NSF channel.
 The color bars indicate the strength of the neutron scattering intensity.   
}
\end{figure}

To conclusively determine the magnetic structure of the $\sqrt{5}\times\sqrt{5}$ superlattice, we 
carried out neutron diffraction experiments on an as-grown 
single crystal of nonsuperconducting Rb$_{0.89}$Fe$_{1.58}$Se$_2$. 
The zero field cooled (ZFC) magnetic susceptibility measurements 
on the sample indicate no bulk superconductivity [Fig. 4(a)].  As discussed in previous work \cite{wbao1,pomjakushin1,fye,wbao2,pomjakushin}, the block AFM structure in Fig. 3(b) can be described by the space group  $I4/m^\prime$, while the AFM structure in Fig. 4(b) has a space
group of $I112^\prime/m^\prime$. Figures 4(e) and 4(f) show the expected AFM Bragg peak intensities  
in the $(H_s,K_s,-1)$ scattering plane for the $I4/m^\prime$ and $I112^\prime/m^\prime$ space 
groups, respectively.  While the intensity of 
the $Q_s=(3,-2,-1)$ reflection is weaker than that of 
the $(2,-3,-1)$ peak
in the block AFM structure, 
$Q_s=(3,-2,-1)$ reflection should be stronger for 
the quaternary collinear AFM structure.
  Comparison of the mapping of the Bragg peaks in the 
$(H_s,K_s,-1)$ scattering plane in Fig. 4(c) with these two models in Figs. 4(e) and 4(f) confirms that the block
AFM structure with space group $I4/m^\prime$ is correct.  The background subtracted raw data for 
$Q_s=(3,-2,-1)$ and $(2,-3,-1)$ Bragg peaks are shown in Fig. 4(d), which again confirm the block 
AFM structure \cite{wbao1}.

\begin{figure}[t]
\includegraphics[scale=.45]{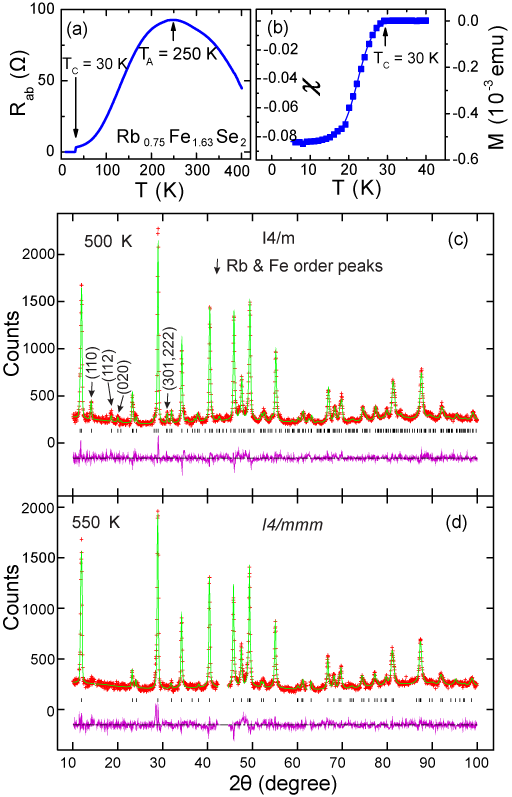}
\caption{(color online)  (a) In-plane resistivity of Rb$_{0.75}$Fe$_{1.63}$Se$_{2}$ indicating $T_c=30$ K.
(b) Susceptibility measurements with 20 mT in-plane field on a 14.2 mg Rb$_{0.75}$Fe$_{1.63}$Se$_{2}$ single crystal confirm the superconducting transition temperature $T_c=30$ K. The left axis is the calculated  volume magnetic susceptibility indicating a $\sim8\%$ superconducting volume fraction. The right axis show magnetic susceptibility in units of emu (1 emu$ = 10^{-3}Am^2$).
(c) Neutron powder diffraction pattern at 500 K showing the refinement results for the sample including 
the $\sqrt{5}\times\sqrt{5}$ superlattice distortion.  The superlattice induced peaks are marked with arrows.
 (d) The neutron refinement results in the paramagnetic
state at 550 K.
}
\end{figure}

To further establish the magnetic nature of the proposed block AFM structure, we have measured 
 all Bragg peaks in Fig. 1(d) by polarized neutrons, where  SF and NSF scattering correspond to pure magnetic and
 pure nuclear scattering, respectively, in the HF configuration.  All measurements
 were done at room temperature allowing easy tilting of the samples to access different Bragg peaks.
Figure 5 summarizes the reciprocal space probed and the raw SF and NSF scattering for different Bragg peaks in the 
tetragonal unit cell notation.  Initially, we 
aligned the sample in the $[H,H,L]_T$ zone as shown in the dashed line of Fig. 5(a). The $[H,H,0]_T$ and
$[0,0,L]_T$ axes are aligned along the lower and upper arc axes of the goniometer, respectively.
By rotating the lower arc of the goniometer by angles $\alpha$ as shown in Fig. 5(a), we can 
access magnetic Bragg peaks 
$Q_T=(1.2,0.6,1)$ and $Q_T=(1.4,1.2,1)$ associated
with the block AFM structure [Fig. 5(a)].  Figures 5(b)-5(e) reveal that the expected magnetic
Bragg peaks only appear in the SF channel, and there are no features in the NSF channel.
Therefore, $Q_T=(1.2,0.6,1)$ and $Q_T=(1.4,1.2,1)$ Bragg reflections are 
are entirely magnetic in origin with no nuclear component.  
To access $Q_T=(0.6,0.2,2)$ and $Q_T=(1.4,0.2,2)$ peaks, we realigned the sample 
to the $[H,0,L]_T$ zone.  Figures 5(f)-5(i) show that $Q_T=(0.6,0.2,2)$ and $Q_T=(1.4,0.2,2)$ peaks 
appear entirely in the NSF channel thus revealing their nuclear origin.  These results conclusively establish the
magnetic nature of the block AFM structure with $I4/m^\prime$ space group for the nonsuperconducting 
Rb$_{0.89}$Fe$_{1.58}$Se$_2$.

\begin{figure}[t]
\includegraphics[scale=.35]{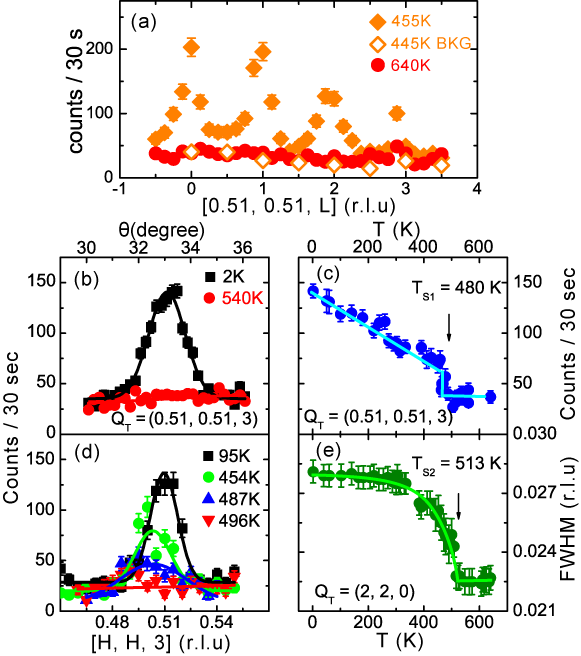}
\caption{ (color online) Triple-axis measurements on a single crystal of Rb$_{0.75}$Fe$_{1.63}$Se$_{2}$ on HB-1A.
(a) Elastic $[0.51,0.51,L]$ scans at different temperature. The scattering is above background scattering at all $L$-values. The background scattering is determined from $[H,H]$ scans. All the peaks are centered at $H\approx 0.5$ (not exactly at 0.5 due possibly to 
a small crystal misalignment on HB-1A).  The strong $L$-modulation disappears at 640 K.
(b) Rocking curves of the (0.51,0.51,3) reflection at 2 K and 540 K.
(c) Temperature dependence of the peak intensity for the (0.51,0.51,3) reflection shows a first-order-like transition above 480 K as marked by the arrow. (d) Temperature dependence of the $[H,H,3]$ scans shows that the $(0.5,0.5)$ lattice
distortion disappears at 495 K, below the tetragonal to $\sqrt{5}\times\sqrt{5}$ superlattice reflection.
(e) Widths of the (2,2,0) nuclear reflection indicates a second-order-like structural transition 
at 513 K. The solid lines are guides to the eye. 
}
\end{figure}

\begin{figure}[h]
\includegraphics[scale=.45]{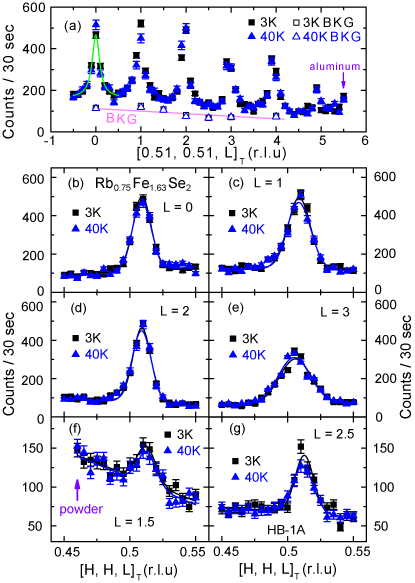}
\caption{(color online) Low-temperature triple-axis measurements on Rb$_{0.75}$Fe$_{1.63}$Se$_{2}$.
(a) The $[0.51,0.51,L]$ scans below and above $Tc$.  
The background are from $[H,H]_T$ scans and solid lines are guides to the eye.
(b-g) Elastic $[H,H,L]_T$ scans at $L=0,1,2,3,1.5,2.5$ below and above $T_c$.  The lattice distortion 
 is not affected by the superconductivity.
}
\end{figure}

Although the block AFM structure for nonsuperconducting Rb$_{0.89}$Fe$_{1.58}$Se$_2$ 
is now firmly established, it is still unclear how
the static AFM order co-exists with superconductivity.  In previous work \cite{wbao1,fye,wbao2}, it was argued that the block AFM order with huge moments in 
$A_y$Fe$_{1.6+x}$Se$_2$ microscopically coexists with superconductivity.  However, recent X-ray diffraction measurements have provided compelling evidence for nanoscale phase separation in K$_{0.8}$Fe$_{1.6}$Se$_2$ \cite{ricc1,ricc2}. To check how superconducting $A_y$Fe$_{1.60+x}$Se$_{2}$ differs from the nonsuperconducting samples, we prepared a single crystal Rb$_{0.75}$Fe$_{1.63}$Se$_{2}$, where transport measurement shows $T_c=30$ K, metallic behavior below 250 K,  and semiconducting characteristics above [Fig. 6(a)].  Although 
magnetic susceptibility confirms the superconducting transition at $T_c=30$ K, we estimate that the superconducting volume fraction in our sample is only around $8\%$ [Fig. 6(b)]. 
To determine the precise crystal lattice structure and atomic compositions, we carried out neutron powder diffraction measurements on BT-1.  Rietveld analysis of the powder diffraction data at 550 K using the $I4/mmm$ space group fits the data well [Fig. 6(d)].  At 500 K, Fe vacancies order into a $\sqrt{5}\times\sqrt{5}$ superlattice structure as shown in Fig. 1(a) \cite{haggstrom} and the powder diffraction pattern can be well described by the space group $I4/m$.

We have searched extensively for structural and magnetic peaks in 
superconducting Rb$_{0.75}$Fe$_{1.63}$Se$_2$.  In addition to confirming 
the $\sqrt{5}\times\sqrt{5}$ AFM peaks at identical positions
as the nonsuperconducting Rb$_{0.89}$Fe$_{1.58}$Se$_2$, we find a set of new peaks at wave vectors
$Q_T\approx(0.5,0.5,L)$ where $L=0,1,2,3,...$.  Along the $c$-axis, these peaks 
are broad and Lorentzian-like, and centered at integer $L$ positions.  They disappear on warming 
from 445 K to  640 K [Fig. 7(a)], suggesting that they are associated with either a magnetic phase transition or structural
lattice distortion not related to the known $\sqrt{5}\times\sqrt{5}$ superlattice structure.
The broad nature of the scattering along the $c$-axis indicates that they are quasi two-dimensional.
Figure 7(b) plots rocking curve scans at wave vector $Q_T=(0.51,0.51,3)$, which again show
 the disappearance of the low-temperature peak at 540 K. To determine the phase transition temperature
 associated with the $Q_T\approx(0.5,0.5,L)$ peaks and compare those with the tetragonal ($I4/mmm$) to 
the $\sqrt{5}\times\sqrt{5}$ superlattice ($I4/m$) transition,
  we carefully measured the intensity of the $(0.51,0.51,3)_T$ peak and the width of the $(2,2,0)_T$ Bragg peak. While the $(0.51,0.51,3)_T$ peak shows an abrupt first-order-like phase transition and disappears above 480 K [Fig. 7(c)], the Bragg peak width on the $(2,2,0)_T$ reflection shows a second-order-like phase transition at 513 K [Fig. 7(e)]. Figure 7(d) shows the $[H,H,3]$ scans at different temperatures, which display no peak at 496 K, thus confirming that the phase transition temperature for the $Q_T\approx(0.5,0.5,L)$ peaks happens at 
 a lower temperature than that of the tetragonal-to-$\sqrt{5}\times\sqrt{5}$ superlattice distortion.

\begin{figure}[h]
\includegraphics[scale=.45]{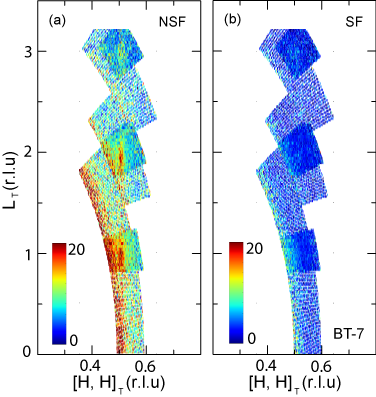}
\caption{(color online) Polarized neutron scattering measurements near the $[0.5,0.5,L]$ Bragg peak for Rb$_{0.75}$Fe$_{1.63}$Se$_{2}$ at room temperature. The exerpimental configuration is the same with that for Fig. 5. 
While the NSF scattering on the left panel shows clear nuclear scattering centered around the  $[0.5,0.5,L]$ positions,
there is no evidence of SF scattering in the same reciprocal region.  This means that there is 
no elastic magnetic scattering at the same wave vector as the iron pnictides such as LaFeAsO \cite{cruz}.
}
\end{figure}

To see if the two-dimensional $Q_T\approx(0.5,0.5,L)$ scattering responds to the formation of the superconductivity, we have carried out 
in-plane and $c$-axis scans below and above $T_c=30$ K on Rb$_{0.75}$Fe$_{1.63}$Se$_2$.  
We find that these peaks do not exhibit any changes across $T_c$ [Fig. 8(a)].
Figure 8(b)-(g) show elastic scans along the $[H,H,L]_T$ direction at $L$-values of $L$=0,1,2,3,1.5,2.5. For all $L$-values, we find peaks centered at $H=0.51$ rlu, confirming the two-dimensional nature
of the scattering. In previous X-ray diffraction experiments on Cs$_y$Fe$_{2-x}$Se$_2$,  Pomjakushin {\it et al.} \cite{pomjakushin} have also found peaks at $Q_T\approx(0.5,0.5,L)$.  This means that 
a portion of the signal we observe at $(0.5,0.5,L)$ must be due to a structural distortion. To determine if there is 
any additional magnetic component in the 
$(0.5,0.5,L)$ scattering, we performed neutron polarization analysis.  Figures 9(a) and 9(b) show mappings of the reciprocal space
in the $[H,H,L]$ zone for the NSF and SF scattering, respectively.  While one can see a clear rod of scattering centered at $L=0,1,2$ 
along the $[0.5,0.5,L]$ direction in the NSF channel, the SF scattering is featureless in the entire probed range.  Since the $(0.5,0.5,L)$ rod-like scattering is not compatible with the $\sqrt{5}\times\sqrt{5}$ superlattice structure, our data suggest that Rb$_{0.75}$Fe$_{1.63}$Se$_2$ is phase separated and exhibits 
two structural transitions, one at 513 K and the second one at 480 K.

\begin{figure}[h]
\includegraphics[scale=.4]{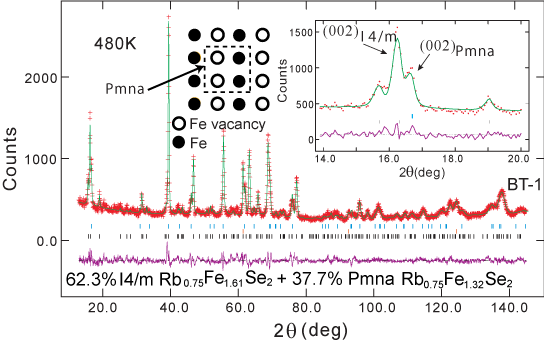}
\caption{(color online) Neutron powder diffraction measurements on Rb$_{0.75}$Fe$_{1.50}$Se$_{2}$ at 480 K. They are fit using a combination of $62.3\%\ I4/m$ and $37.7\%\ Pmna$ phases. The purple trace indicates the intensity difference between the observed (green solid line) and calculated (red crosses) structures. The left inset is a schematic of the $Pmna$ phase which can induce nuclear peaks at $[m\pm 0.5,n\pm 0.5,L]_T$ ($m,n=0,\pm1, \pm2, \pm 3...$).
}
\end{figure}

In an attempt to determine the lattice structure associated with the $[0.5,0.5,L]$ reflections, we carried out detailed Rietveld analysis on the neutron powder diffraction pattern.  
Since the $[0.5,0.5,L]$ reflections are much weaker in the powder pattern, we were able to fit the powder diffraction pattern with a combination of $62.3$\% $\sqrt{5}\times\sqrt{5}$ superlattice ($I4/m$) and  
Rb$_{0.75}$Fe$_{1.32}$Se$_2$
Fe-vacancy model in the inset of Fig. 10 ($Pmna$) \cite{wbao2}.  However, such an Fe-vacancy model with $Pmna$ space
group will not be able to explain the quasi two-dimensional rod scattering we observe in the  
triple-axis measurements.  Therefore, it remains unclear what crystalline lattice distortion gives rise to the
observed superlattice reflections, although we know such scattering enlarges the nuclear unit cell by 
$\sqrt{2}\times \sqrt{2}$.  We note that a recent X-ray study on superconducting Rb$_y$Fe$_{1.6+x}$Se$_2$ samples
also found a $\sqrt{2}\times \sqrt{2}$ superlattice structure \cite{tsurkam}.

\section{IV. Discussions and Conclusions}

Using single crystal neutron diffraction and neutron polarization analysis, we have confirmed that the block AFM structure in Fig. 1 is the only possible magnetic structure for insulating AFM Rb$_{0.89}$Fe$_{1.58}$Se$_2$ \cite{wbao1}.  Although we have also found a similar AFM structure for the superconducting Rb$_{0.75}$Fe$_{1.63}$Se$_2$, careful analysis of the diffraction spectra reveals another structural phase transition associated with enlarged unit cell for superconducting
Rb$_{0.75}$Fe$_{1.63}$Se$_2$.  In previous ARPES measurements on superconducting $A_y$Fe$_{1.6+x}$Se$_2$ \cite{yzhang,xpwang,xjzhou}, different groups have reached the same conclusion concerning the electron-like Fermi surfaces  at $M(\pi,0)/(0,\pi)$ points.  However, there have been debates concerning the origin of the observed weak electron pockets near the $\Gamma(0,0)$ point \cite{yzhang,xjzhou}.  In principle, the electron pockets near the $\Gamma$ point can arise from 
band folding if there exists a $(0.5,0.5)$ structural or magnetic phase transition \cite{xjzhou}.  Our observation of  
the quasi two-dimensional $(0.5,0.5,L)$ superlattice reflections suggests that the observed electron Fermi surfaces near the
$\Gamma(0,0)$ point may indeed be due to band folding instead of a surface state.  Since the block AFM structure in insulating Rb$_{0.89}$Fe$_{1.58}$Se$_2$ cannot arise from Fermi surface nesting, we speculate that the $\sqrt{2}\times \sqrt{2}$ lattice distortions in superconducting Rb$_{0.75}$Fe$_{1.63}$Se$_2$ may be associated with the metallic portion of the sample.  In this picture, the superconducting phase in $A_y$Fe$_{1.6+x}$Se$_2$ may be mesoscopically phase separated from the nonsuperconducting phase,
where superconductivity and AFM order are intertwined in a very short length scale and
live in separate regions.  Theoretically, it has been suggested that the $A$Fe$_{1.5}$Se$_2$ phase is 
a semiconductor with a low energy band gap \cite{txiang}.  So with electron or hole doping, such a phase would
become nonmagnetic and superconducting.
 Although we have no direct proof that the superconducting portion of the sample is associated with the $(0.5,0.5,L)$ superlattice distortion, systematic neutron scattering 
measurements are currently underway to investigate the relationship of such phase to the block 
AFM order and superconductivity.

\section{V. Acknowledgements}   
We are grateful to Jiangping Hu and Tao Xiang for helpful discussions. 
Work at IOP is supported by the Ministry of Science 
and Technology of China (973 Project No. 2010CB833102, 
2010CB923002, 2011CBA00110), and Chinese Academy of Sciences. 
The single crystal growth and neutron scattering effort at UT is supported by U.S. DOE BES under Grant No. DE-FG02-05ER46202 (P.D.).
Work at ORNL neutron scattering facilities are supported by 
the Scientific User Facilities Division, Office of Basic Energy Sciences, U.S. Department of Energy. 
Part of the work at UT is also supported by the U.S. NSF-OISE-0968226 (P. D.).
The work at ZJU is supported by 
by Natural Science Foundation of China (Grant No. 10974175), the Ministry of Science 
and Technology of China (973 Project No. 2011CBA00103).


\end{document}